# Do specific ion effects influence the physical chemistry of aqueous graphene-based supercapacitors? Perspectives from multiscale QMMD simulations.

Joshua D. Elliott[1,2], Mara Chiricotto[1], Alessandro Troisi[3] and Paola Carbone[1]

[1]Department of Chemical Engineering and Analytical Science, University of Manchester, Manchester M13 9PL, United Kingdom
[2]Diamond Light Source, Diamond House, Harwell Science and Innovation Park, Didcot, Oxfordshire OX11 0DE, United Kingdom
[3]Department of Chemistry, University of Liverpool L69 7ZD, United Kingdom

## Abstract

Whether or not specific ion effects determine the charge storage properties of aqueous graphene and graphite-based supercapacitors remains a highly debated topic. In this work we present a multiscale quantum mechanics – classical molecular dynamics (QMMD) investigation of aqueous mono- and divalent salt electrolytes in contact with fully polarizable charged graphene sheets. By computing both the electrochemical double layer (EDL) and quantum capacitance we observe a constant electrode specific capacitance with cationic radii and charge. Counterintuitively, we determine that a switch in the cation adsorption mechanism from inner to outer Helmholtz layers leads to negligible changes to the EDL capacitance, this appears to be due to the robust electronic structure of the graphene electrodes. However, the ability of ions (such as $K^+$) with a relatively low hydration free energy to penetrate the inner Helmholtz plane and adsorb directly on the electrode surface is found to slow their diffusion parallel to the interface. Ions in the outer Helmholtz layer are found to have higher diffusivity at the surface due to their position in ion channels between water layers. Our results show that surface effects such as the surface polarization and the partial dehydration and local structuring of ions on the surface underpin the behaviour of cations at the interface and add a vital new perspective on trends in ion mobilities seen under confinement.

1. Introduction

Aqueous carbon-based supercapacitors represent a chemically stable, low-cost and non-toxic solution to low charge storage capacities and slow charge/discharge rates in conventional capacitors and batteries. Graphene,[1–6] graphite and carbon nanotube[7,8] electrodes each offer comparatively high surface areas that increases the number of ionic species that can be stored electrostatically within the electrochemical double layer. The reversible, non-Faradic physisorption of ions at the electrode surfaces mediates a rapid release of stored charge and therefore drives the function the supercapacitor. The challenge that remains is to identify and fine tune exact electrode-electrolyte compositions that simultaneously maximise the density of surface ion adsorption and minimizes the charge/discharge rate.

Critical to the continued development of aqueous carbon-based supercapacitors is an understanding of whether or not ion specificity plays a role in the amount of charge stored at the interface, and if so: to what extent?[9] Currently, experimental measurements in this area are conflicted, with mixed reports on the behaviour of different ions. The capacitance obtained from graphene films (synthesised from graphene oxide) in contact with LiCl, NaCl and KCl supports the idea that water as a polar solvent inhibits any specific ion effects.[10] On the contrary, in a separate investigation of the capacitance of graphite electrodes,[11] a survey of group 1 chloride salts found that, while specific ion effects could not be detected for the edge-oriented graphite, a clear increase in the capacitance was measured with increasing ionic radii. These specific ions effects were linked to the decreasing free energies of hydration as the cations get larger.[12]

Atomistic simulations based on either fully classical models or electronic structure theory provide an atom-scale resolution of the structuring and static- and dynamical properties of the electrode-electrolyte interface. These simulations increase in computational complexity from classical molecular dynamics (MD) that uses force fields to model the interactions between atoms to first principles molecular dynamics (FPMD). Specifically in the investigation of the charge storage mechanism in aqueous supercapacitors, the non-bonded electrostatic interactions brought about by ion (or water) induced polarization of the electrode surface can be critical in reproducing experimental observations.[13–15] These interactions are explicitly included in FPMD, which enables investigation of the properties of many different aqueous electrode interfaces.[16–19] However, large scale models that capture the entropic effects, known to be important for interfacial properties, require simulations of several thousands of atoms for tens of nanoseconds, which for FPMD methods are prohibitively expensive. This means that a balanced accuracy-computational viability trade-off must be established.

Parameterization of the MD non-bonded potential to account for polarized ion-surface interactions facilitates simulations on the necessary time and length scales,[20] with the caveat that the cumbersome parameterization must be carried out for each and every ionic species, solvent model, surface morphology and surface charge density. In addition, the static nature of the polarization contained in Lennard-Jones potential can overpredict binding energies by more than 10 kJ mol$^{-1}$.[15] Alternatively, a description of the dynamical surface polarization can be introduced into the MD force field by tethering a dummy charge to the atom via a harmonic spring.[21] Similarly, this type of polarizable force field requires a careful parameterization for scenarios where the surface electrode is charged, but also limits the polarizability of the surface to localized spatial regions in the vicinity of each atom and therefore potentially misses long-ranged polarization that results from redistribution of electronic charge density. For strictly metallic surfaces, where the potential in the electrode is fixed, the constant potential method modulates the electrode atom partial charges in response to the specific interfacial electrolyte configuration.[22–25] The validity of this approach for semimetallic electrode surfaces was recently called into question.[14] On the other hand, quantum mechanical (QM) simulations of the ion-surface interactions can be carried out in the presence of an implicit solvent model.[26–28] Removing explicit water molecules drastically reduces the computational cost of the QM simulations and in addition, the treatment of ion and surface electron densities also introduces charge-transfer (CT) interactions, that can contribute to the stabilisation of ions at the surface,[29] to the model. Yet, this comes at the expense of a lack of both the interface dynamics and the explicit ion-hydration shell that contribute to specific adsorption behaviour.

Leveraging the explicit polarization of QM approaches and the rapid simulation of large length and time scales in the classical MD regimes, we previously introduced an iterative QMMD scheme that embeds the dynamical electronic structure of the electrode within a classical force field.[14] This method retains a description of the long-ranged polarization of the surface electrode irrespective of its metallicity, while allowing for nanometre and nanosecond simulations. However, fully classical treatment of the electrolyte phase necessarily means that CT interactions that occur on the Angstrom scale (when surface atoms and ions are in contact) are not included. Our approach diverges from typical QM/MM schemes where the focal point of the calculation is the region that is treated quantum mechanically. In such situations, the target of the simulation is evolution of the electronic structure of a surface or molecule in response to a dynamic but classically treated solvent. The time scales for processes involved for such a simulation are on the ns scale. To address the evolution of the classical electrolyte which can take places over tens to hundreds of ns, in response to the quantum mechanical polarization, we developed the QMMD method that focuses on the evolution of the classical system, leveraging as efficient as possible QM simulations.

In this work, we deploy our QMMD framework to model the capacitance of graphene electrodes as depicted in Figure 1 at three fixed surface charge densities: the neutral case and charged negatively and positively with ±0.061 Cm$^{-2}$. We consider 1.0 M solutions of LiCl, NaCl, KCl, MgCl$_2$ and CaCl$_2$ recently parameterized for the TIP4P/2005 water model using scaled ionic charges.[30] Further simulation details are provided in the following section. We uncover a switch in the monovalent cationic-adsorption mechanism as the ionic radius increases, which helps to resolve outstanding controversies in specific ion effects at the graphene interface.

2. Methods and Systems

In this work we consider graphene electrodes in contact with a series of aqueous electrolyte solutions at 1.0 M concentration, and three different surface charge densities: $\sigma^+ = 0.061 \text{ Cm}^{-2}$ ($0.382 \ e \ nm^{-2}$), $\sigma^0 = 0.000 \text{ Cm}^{-2}$ ($0.000 \ e \ nm^{-2}$) and $\sigma^- = -0.061 \text{ Cm}^{-2}$ ($-0.382 \ e \ nm^{-2}$). We model a $3 \times 3$ nm$^2$ infinite graphene sheet comprised of 336 C atoms in periodic boundary conditions. Prior to the QMMD simulations the structure of the graphene has been optimized at the PBE-DFT[31] level using the Quantum Espresso software suite[32,33] and Optimized Norm Conserving Vanderbilt pseudopotentials.[34] The final lattice constant is 0.247 nm, which yields a C-C bond length of 0.143 nm. In contact with the graphene sheet are 8 nm slabs of LiCl, NaCl, KCl, MgCl$_2$ and CaCl$_2$ solutions, a further 8 nm of vacuum separates periodic replicas, the simulation cell has dimensions $2.96880 \times 2.99957 \times 16.0000$ nm$^3$ and is depicted in Figure 1. We have elected to model the capacitor systems using a single electrode configuration; a classical supercapacitor with parallel electrodes presents challenges associated with implementation of two quantum mechanical surfaces compared with the modest advantages provided by having both electrodes present within the same simulation. On the other hand, the single electrode configuration naturally constrains the total charge of the electrode during the SCC-DFTB determination of the surface charges. This prevents spurious charging/discharging that can occur via charge hopping between electrodes in a 2-electrode setup. While the electrode is charged neutral the electrolyte composition is as follows: for monovalent cations, 2214 water molecules, 41 cations, and 41 anions. For divalent cations, 2169 water molecules, 43 cations, and 86 anions. At negative electrode charges to maintain overall electroneutrality we modulate the number of cations in the system; our electrolytes contain 45 cations.

Conversely, at a positively charged electrode we modulate one or both ions: monovalent electrolytes have 45 anions, divalent electrolytes have 42 cations and 88 anions.

The simulations presented here have been carried out using a QMMD-scheme introduced in earlier our work.[14] This is an iterative process that couples quantum mechanical self-consistent charge (SCC) density functional Tight-Binding (DFTB) calculations of the electrode electronic structure to classical molecular dynamics (MD) trajectories of the electrode immersed in a electrolyte. Contrary to more standard QMMM recipes that probe the dynamical evolution of a quantum system embedded in a classical electrostatic background, the objective of our approach is to capture the structuring and dynamical evolution of the interface due to the time dependent electrolyte-induced polarization of the surface. The QMMD scheme proceeds via 4 steps: **(i)** at a given time, $t$, the electrolyte atoms, typically those belonging to solvent molecules and dissolved ions, are converted to a set of point charges using their Cartesian coordinates and force field partial charges. **(ii)** The electronic structure of the surface is then computing using this set of point charges as a background electrostatic potential. **(iii)** Post processing of the surface electronic structure furnishes the set of atomic charges via Mulliken Population analysis. **(iv)** Finally, the surface atomic charges are used to update the graphene partial charges inside the classical force-field and a short dynamical trajectory is carried out. It should be stressed, in this approach only the surface atoms are polarizable and the ionic and solvent atoms have fixed partial charges according to their respective classical force fields. Iteratively looping over these four steps drives the classical dynamics by means of Newtons equations including feedback from the redistribution of surface charge density captured by the DFTB. In this work, unless otherwise stated QMMD simulations have been carried out for 60 ns with the coupling time between QM and MD approaches was set to 5 ps. We previously investigated the effect of different coupling times on the computed surface partial charges for this type of system,[14] finding that 5 ps represents a good accuracy-viability trade off, when compared with an update every MD time step. Of the total simulation time 10 ns is reserved for equilibration.

DFTB calculations are carried out using the DFTB+ software package,[35] with empirical parameters used to describe C-C interactions taken from the mio-1-1 set.[36] Our previous work determined that the computed atomic charges for the systems studied here are fairly insensitive to strict convergence parameters and additionally are still in good agreement with full DFT calculations carried out at the hybrid B3LYP level of theory.[14] As such, to expedite the QM step we perform $\mathbf{\Gamma}$-point only simulations with a Fermi temperature of $1 \times 10^{-6}$ K for filling the electronic states. The convergence threshold on the self-consistent charge optimization is set to $1 \times 10^{-2}$ Hartree.

Classical MD simulations are performed in the NVT ensemble using GROMACS, version 2018.4.[37,38] We employ the Nosé-Hoover thermostat fixing the temperature at 300 K, with a coupling $\tau_1 = 0.1$ ps. The particle mesh Ewald approach is used to treat long-ranged electrostatic interactions, the cut-off between long and short ranged computation is set to 1.4 nm. Pairwise non-bonded interactions are described by a Lennard Jones 12-6 potential smoothly truncated at 1.2 nm using a switch function with onset from 1.0 nm. Water in our simulations is described using the rigid TIP4P-2005 model,[39] rigidity is maintained via the SETTLE algorithm.[40] Mono and divalent ion parameters are taken from the Madrid-2019 force field.[30] We use the Amber force-field for carbon electrode atoms in the graphene sheets, which are frozen throughout the trajectory. Non-bonded ion-carbon and oxygen-carbon interaction potentials are computed using Lorentz-Berthelot combining rules.

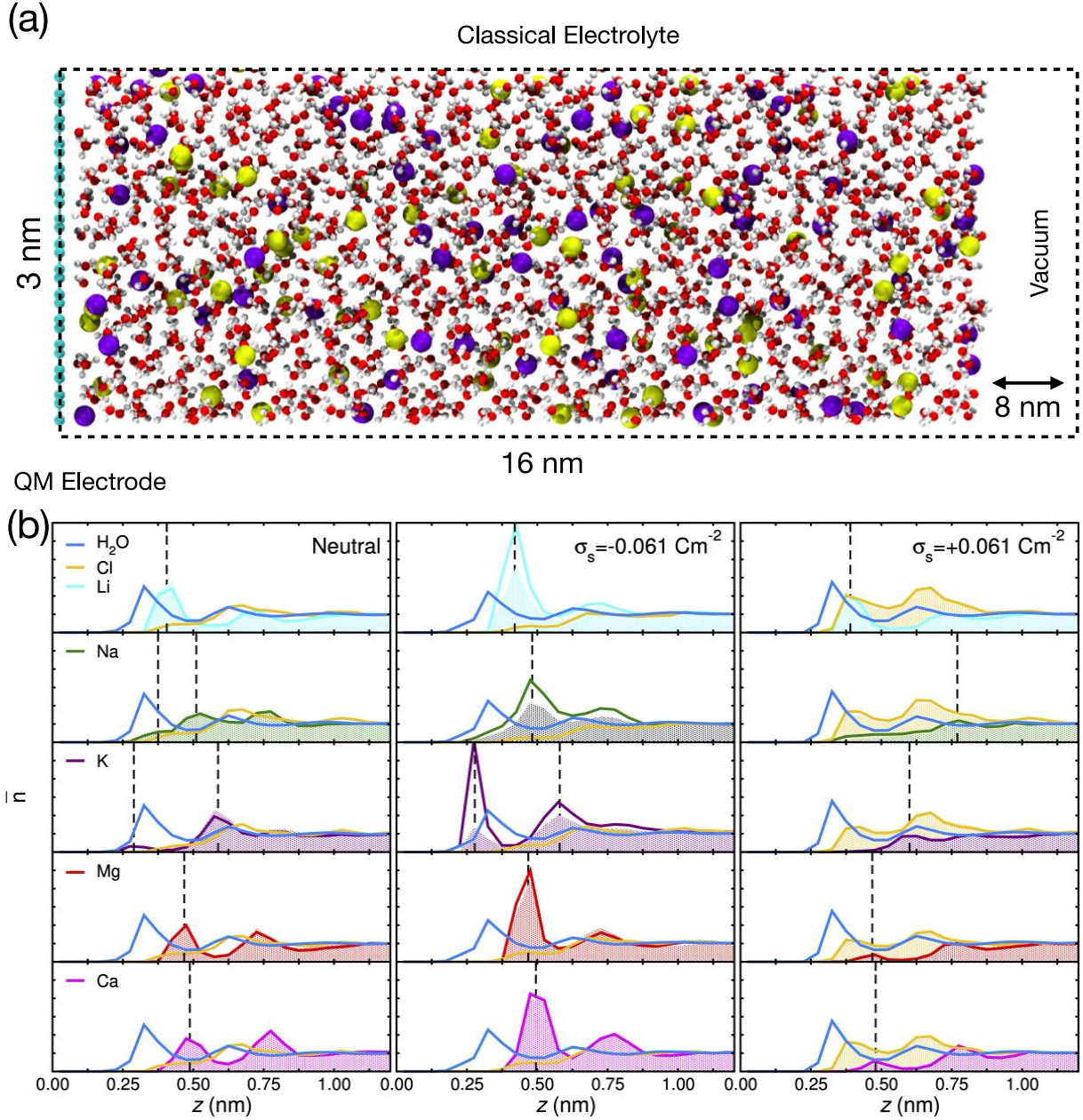

**Figure 1:** (a) *Representative simulation box for the graphene-aqueous electrolyte interfaces considered in this work. The two-dimensional graphene electrode (cyan), approximately $3 \times 3$ nm$^2$ in dimension, is placed in contact with 8 nm slabs of electrolyte (cations: purple, anions: yellow, O: red, H: white) with a further 8 nm vacuum buffer to prevent interactions between periodic images. (b) Plot of the bulk normalized densities along the surface normal for each of the different electrolytes considered for a neutral (left) and negative (centre) and positively (right) charged electrode. Vertical dashed lines denote the positions of the first cation adsorption peaks. Solid lines represent the density profiles of the QMMD simulations, shaded regions represent density profiles of the cations and anions in fully classical simulations.*

### 3. Results and Discussion

*3.1 Structure of graphene-electrolyte interface*

We first examine the structure of the various charge neutral graphene-electrolyte interfaces at 1M concentration, including the effects of the quantum mechanically polarized surface. The number density profile along the axis normal to the surface describes the coarse structuring of the electrochemical double layer at the interface. To extract information from the density profile we introduce the bulk-normalized number density, $\bar{n}(z)$, which is scaled so that that density for each given species in the bulk region is unity. The $\bar{n}(z)$ are plotted in Figure 1.

Across all the different electrolyte solutions, the water (blue, Figure 1) and Cl anion (yellow, Figure 1) density profiles do not change appreciably; the respective first adsorption peaks are located at 0.33 and 0.65 nm. The interfacial water adopts the classic structure, having two separate peaks; the placement of these peaks is consistent with previous investigations of graphene and the TIP4P-2005 water model.[41,42] We note that the Cl anion is repelled from the neutral surface, a finding consistent with conclusions of Misra *et al* that strongly hydrated kosmotropic anions are generally repelled by graphene.[15]

In the case of the monovalent cations, the $\bar{n}(z)$ of the smaller Li and Na indicate that these ions reside in the outer Helmholtz plane, likely due to their respective high free hydration energies[43] and kosmotropic behaviour preventing partial dehydration of the ion first hydration sphere, this is discussed in greater detail below. In the case of Li, the first adsorption peak is found at 0.41 nm and the second at 0.72 nm, whereas Na has a minor adsorption peak at 0.37 nm, a major adsorption peak at 0.51 nm and a second adsorption peak at 0.75 nm. Each of these peaks is further away from the electrode than first water layer, which is consistent with our previous finding for Na (0.49/0.73 nm[14]) using the QMMD approach,[14] which incidentally used a different set of (non-scaled partial charge) classical parameters developed by Joung and Cheatham for the classical force field.[44–46]

Like the Na ion, the larger K cation (purple, Figure 1) also has minor and major adsorption peaks, these are located at 0.29 nm and 0.58 nm, with a second adsorption peak at 0.79 nm. It is noteworthy that the minor peak is closer to the graphene surface than the first water peak indicating that the K ion can penetrate the inner Helmholtz plane. This is at variance with the finding of Pykal *et al*, who investigated aqueous KF- and KI-graphene interfaces using a polarisable force field model, and always observed K in the outer Helmholtz or diffuse layer.[47] In their work, it was deemed that the K ion adsorption behaviour was linked to the adsorption behaviour of the anion, with iodide able to displace K at the surface and fluoride able to draw the K cation into the bulk. It is therefore conceivable that the position of the Cl ion in the outer Helmholtz layer facilitates adsorption of K at the interface. We tentatively suggest that this could also be linked to the ability of our model to delocalize surface charge density to the adsorption site, an effect not captured with the Drude oscillator polarizable force field,[47] as well as to the relative hydration free energy of the K ion across different models and force fields. However, we caveat this conclusion by pointing out that simulations of KSO4[48] by the constant potential method[22,49] (which is also able to the delocalise charge) results in minimal change to outer Helmholtz EDL structuring of Li-, Na- and KSO4. We highlight that the large K cation is generally considered to be chaotropic and therefore the adsorption of K in the inner Helmholtz layer in our work is in further agreement with similar conclusions drawn for anions,[15] our findings suggest that chaotropic cations are also attracted to the neutral graphene surface. While there is diversity in the profiles of the monovalent cations, the two divalent cations Mg (red, Figure 1) and Ca (magenta, Figure 1) have highly similar density profiles: the first and second adsorption peaks are located at 0.47 & 0.73 nm and 0.49 & 0.78 nm respectively, in the outer Helmholtz and diffuse layers.

In addition to the examination of the behaviour of the system setup with a charge neutral electrode, we also computed the $\bar{n}(z)$ for 1.0 M electrolyte concentrations in contact with electrodes charged positively and negatively, these are also presented in Figure 1. In order to maintain overall electroneutrality, the electrode surface charge density has been set to ±0.061 C m$^{-2}$ and the number of cations and anions in the electrolyte solution modulated to balance the electrode charge.

Our results show that upon charging, only the K ion enters the inner Helmholtz layer and directly adsorbs on the graphene surface. When the graphene electrode is charged negatively the positions of main peaks in the $\bar{n}(z)$ profile relative to the graphene electrode remain the same. However, the respective peak intensities, specifically those at the interface, vary drastically. At 1.0 M concentration, the first and second adsorption peaks for Li are again found at 0.4 and 0.7 nm from the graphene surface. As with the neutral electrode, these are still in the outer Helmholtz plane since the first water peak is at 0.3 nm, but now with respective intensities 5 and 1.5 times larger than the bulk density. The same effect is seen when we consider Na, the first and second adsorption peaks are observed at 0.5 and 0.7 nm respectively, the peak intensities are 3 and 1.5 times larger than the bulk density. On a technical note, the fact that the cation peaks do not change position upon negative charging of the electrode indicates that the in the trade-off between the repulsive part of the Lennard-Jones interaction potential and the attractive surface polarization-augmented Coulomb potential, the Lennard-Jones interaction dominates. This can be explained from the results of Misra *et al*, who identified using a bespoke Drude oscillator model that (for the neutral case) interference between water and ion electric fields can effectively screen up to 85 % of the surface ion attractions.[15] While here we consider the negatively charged electrode, we still expect some degree of this screening to take place. The small difference in the positions of the first adsorption peak of the Li and Na (Na 0.1 nm further from the surface) is likely due to the larger ionic radius and solvation sphere of the Na ion. The result is a slightly weaker attractive Coulomb interaction with the charged surface and therefore reduced adsorption intensity; averaged over 15 ns of QMMD simulation the short-ranged Coulomb interaction between the anode and ion is -28.4 kJ/mol in the case of Li and -20.9 kJ/mol for Na. As shown in Figure 1, for K the intensity of the first peak at 0.29 nm, which is within the inner Helmholtz plane, is six times greater than the bulk concentration indicating a significant build-up of ions in direct contact with the negatively charged electrode. Given the proximity of the ion to the surface, as well as the overall negative charge of the electrode, we observe that the K ion has a stronger Coulombic attraction to the polarizable surface -74.6 kJ/mol, approximately three times greater than the other monovalent ions in the outer Helmholtz layer, which suggests a more minor role for the screening described by Misra and Blankstein.

Turning to the divalent ions, the first and second adsorption peaks are again found at 0.47 & 0.73 nm for Mg, and 0.50 & 0.76 nm for Ca. The intensities of the first adsorption peaks are five and four times the bulk ion density respectively, which further indicates that there is little in the density profile to discriminate between these two ions. The Coulombic attraction between the divalent ions and the electrode is expectedly larger than the monovalent ions Li and Na, Mg -40.5 kJ mol$^{-1}$ Ca -30.4 kJ mol$^{-1}$, this is again due to their rigid solvation shell, similar proximity to the surface and larger ionic charge.

Except for the 1M LiCl solution, charging the electrode positive has the same effect on all the systems considered. The cation first adsorption peaks closest to the electrode are dramatically reduced, leaving just a small number of cations in the double layer, in the case of K we in fact see almost zero cations, this effect is due to the trade off between positively charged ions being repelled by the surface yet attracted by the nonlocal forces such as dispersion described by the classical force field. On the other hand, Li is only minimally affected by the presence of the positively charged electrode, which could be due to its specific carbon-lithium non-bonded interaction. Instead, we observe the outer Helmholtz layer adsorption of the Cl anions at approximately 0.4 nm from the surface with an intensity two times the bulk density.

In order to determine which properties are the driving force for the K ion to penetrate the inner Helmholtz layer, in contrast to the other cations considered, we performed a series of 6 ns simulations with a negative charging of the electrode. In each simulation one cation was frozen either in the bulk solution at a perpendicular distance ($d_{Bulk}$) from the surface or close to the surface at a distance ($d_{Surface}$); the surface absorption height ($d_{Surface}$) was chosen according to the first adsorption peak in the density profile as reported in Table 1, in this way, we probe the configuration of the ion as it is absorbed on the surface. Figure 2 reports the radial distribution functions (cation-oxygen) and total number of coordinated water molecules within a given radius in bulk (a) and adsorbed on the surface (b). Within the first coordination sphere, which is 0.35 nm, the difference between the bulk and surface is minimal (< 0.1 water molecules) for all ions except K. In fact, for Li, Na, Mg and Ca the radial distribution function suggests that all cation-oxygen distances in the first solvation shell fall within 0.5 Å of each other (within the range 0.25-0.3 nm). This helps to explain why these ions all present with first adsorption peaks at similar distances from the electrode surface (Figure 1b); i.e. dictated by the closest approach of water in the first solvation shell to the surface. K on the other hand has on average 1.5 less coordinated water when it is adsorbed at the surface. These results are highlighted in Figures 2c and 2d, which show the coordination of a Na and K ion respectively at the charged surface, and highlight that the loss of water molecules from the K solvation shell is due to direct adsorption and not due to other surface induced configurations (partially solvated ion pairs or lower coordinated geometries). For Na, the closest 5 water molecules make up the first coordination shell, of which two clearly reside in between the ion and the surface. In contrast the K ion is directly adsorbed on the graphene surface with no intermediate water molecule. We note that even in the bulk K does not have a stable first solvation shell; unlike each of the other ions the K radial distribution function has a broad peak with an elongated tail which signifies that the final molecule in the coordination sphere is weakly bound at a moderately large distance from the ion. Consequently, there is a comparatively low energy requirement to remove such a molecule from the solvation shell with respect to the other ions investigated that permits the dehydration we observe at the surface. Behaviour which is consistent with results from investigations of K ion migration through biological channels.[50]

**Table 1:** *Vertical distance between the electrode and the frozen cation in the bulk electrode and at the interface. Surface ions have been fixed at the vertical coordinate corresponding to the first peak in the density profile.*

| Electrolyte | $d_{Bulk}$ (nm) | $d_{Surface}$ (nm) |
|---|---|---|
| LiCl | 3.991 | 0.414 |
| NaCl | 3.960 | 0.393 |
| KCl | 4.328 | 0.280 |
| $MgCl_2$ | 4.256 | 0.474 |
| $CaCl_2$ | 4.207 | 0.487 |

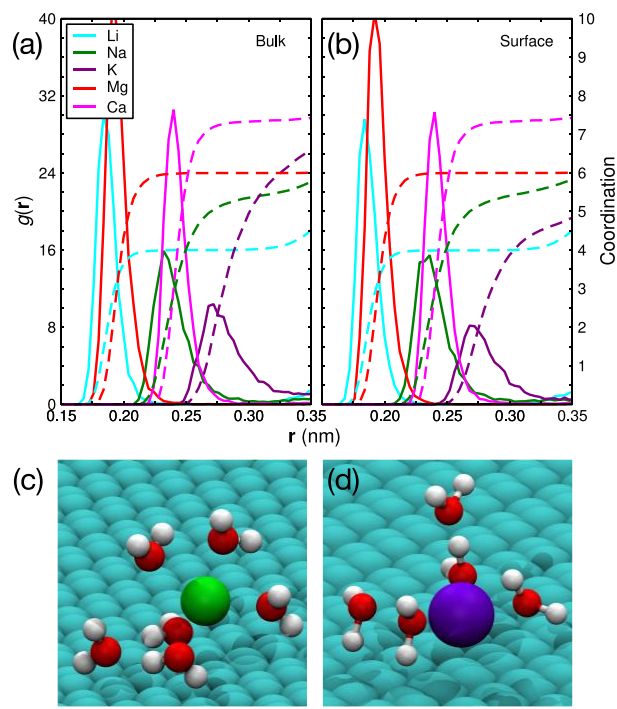

**Figure 2:** *Computed radial distribution functions (ion-oxygen) (solid lines) and integrated radial distribution functions (dashed lines) of frozen cations in the bulk solution (a) and at the surface (b) for the various 1M concentration electrolytes considered in this work. Representative configurations of the Na (c) and K (d) ions frozen at the interface.*

The partial dehydration of the K ion compared with the other ions we studied derives from the classical Lennard-Jones potential, while the quantum mechanical calculations return the correct electrostatic potential for all of the ions. The resulting direct exposure of the K ion to the surface leads to an induced polarization of the C atoms close to the ion resulting in strongly negative attractive Coulomb interactions. We previously investigated the radial charging of a graphene flake in response to the surface adsorption of positive and negative ions.[14] The surface charge redistribution induced by the proximity of an ion is sizable and nonlocal, affecting carbon atoms (in rings surrounding the adsorption site) as far as three bonds suggesting that atomic-centred polarizable models can fall short in capturing the full physics of the physisorption process. In Figure 3a, for a single example configuration, we plot the surface charges in two dimensions and mark the positions of K ions directly adsorbed on to the graphene with crosses circled blue. In this example the surface Mulliken charges are significantly more negative in the surface region local to the positive ion.

To understand the extent to which this is a general property, we also plot in Figure 3b all surface charges as a function of the absolute distance from each adsorbed K ion calculated for 15 ns of the full 60 ns trajectory. It's worthwhile to highlight we only included absolute distances up to 5 Å to avoid situations where the K ions have little to no influence on the computed partial charge. The computed Mulliken charges decrease linearly as the ions gets closer to the surface; between 0.25 and 0.35 nm the polarization of the C atoms is very strongly influenced by the ion, this corresponds to nearest and immediately bonded atoms in the surface. Beyond 0.35 nm, the computed Mulliken charges average out to the homogenous fixed charge value through interactions with other ions or water molecule dipoles. Our results demonstrate the importance of the local ion induced polarization in the stabilization of the directly adsorbed configuration and this clear trend and distance dependence opens up for the

possibility of an analytical description of the polarization energy to further expedite QM simulations.[51]

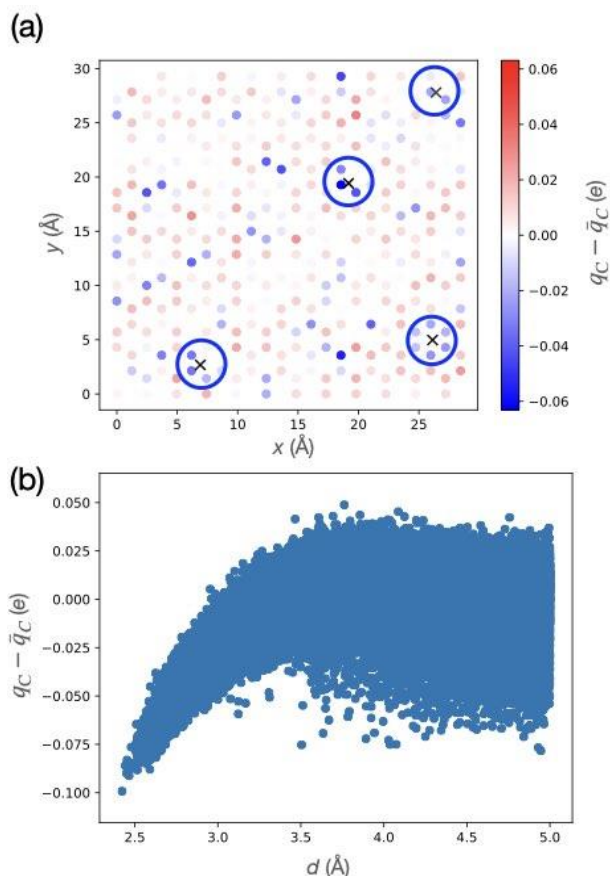

**Figure 3:** *(a) Representative plot of the computed Mulliken charges on the graphene sheet in contact with a 1M solution of KCl and charged with 3.4 e. Circled X's mark the coordinates of K ions directly adsorbed on the surface. (b) A plot of the computed Mulliken charge as a function of the distance from each K ion adsorbed on the surface over 15 ns of QMMD trajectory. For illustrative purposes, all computed charges are normalized to the homogenously averaged surface charge -3.4/336 e (where $-3.4e$ is the total charge of the graphene sheet made of 336 carbon atoms).*

By comparing the results of our QMMD approach with purely classical (MD) simulations of the same systems a further piece of information we can elicit is the role (and importance) of the dynamic surface polarizability on the interfacial behaviour of the water and ions. To this end, we also performed analogous simulations of each of the graphene electrolyte interfaces with homogenous fixed surfaces charges totalling 0.00 and ±0.061 Cm$^{-2}$ respectively. For comparison, the bulk normalized number densities of the absorbing ions in the MD simulations are plotted as shaded regions in Figure 1b.

For the interfaces at the neutral electrode, the first thing to note is that the density profiles for all the systems that do not directly absorb on the surface do not appreciably change between the QMMD and MD simulations. Only in the case of the K ion, which presents with a minor absorption peak at 0.29 nm at the QMMD level corresponding to the absorption configuration with 1 water molecule removed from the first solvation shell. When treated with purely classical MD the minor absorption peak disappears, and the main peak at 0.58 nm increases in intensity. This suggests that the redistribution of charges on the surface, which enables a local accumulation of negative charge (on the overall neutral electrode) is a key factor in overcoming the thermodynamic drive towards dehydration and adsorption of K ions within the Helmholtz layer.

As can be seen from Figure 1b, the effect of the polarization is amplified when the electrode is charged negatively. For all the monovalent cations considered, the intensity of the cation adsorption peaks is significantly greater (with respect to the bulk concentration) in the QMMD simulations compared with the purely classical MD simulations. The most interesting case is the K ion: The first adsorption peak of the K ions at the surface of the negatively charged electrode is 5 times larger in the QMMD case compared with the MD case, while the second adsorption peak is also more intense. In addition, in the MD simulation there is a higher concentration of cations in the second adsorbed layer than the first. What becomes apparent from these simulations is that the charged surface provides sufficient stabilization of the adsorbed configuration through the Coulomb potential. It turns out that the dynamic polarizability then has a very large effect on the concentration of ions that can be absorbed. Again, this is due to the QMMD approach being able to localize negative charge close to the adsorption site further stabilizing the dehydrated adsorbed configuration. Somewhat counterintuitively, divalent ions are largely insensitive to the polarization effects of the surface. This could be due to the presence of more water molecules in their first solvation shell (Figure 2a), effectively screening the ion-electrode interaction.

*3.2 Electrochemical Double Layer Capacitance*

We turn now to the electrochemical double layer integral capacitance, which is obtained from the system number density profile along the surface normal, $n(z)$. First, the charge density, $\rho(z)$, is calculated from the atom-resolved number densities multiplied by their partial charges from the classical force field. This includes all charged atoms (appropriately scaled[30]), ions and electrodes. Poisson's equation relates the curvature of the electrostatic potential, $\Phi$, to the charge density, which along one dimension reads,

$$\frac{\partial^2 \Phi}{\partial z^2} = -\frac{\rho(z)}{\varepsilon_0 \varepsilon_r}, \quad\quad\quad 1$$

where $\varepsilon_0$ is the vacuum permittivity constant and $\varepsilon_r$ is the specific electrolyte permittivity. Integration of Poisson's equation yields the electrostatic potential,

$$\Phi(z) = -\frac{1}{\varepsilon_0 \varepsilon_r} \int_0^z dz' \, (z - z') \rho(z') \quad\quad\quad 2$$

Equation 2 is integrated from a position in the bulk electrolyte where the net charge density is zero, towards the quantum mechanically treated electrode. The bulk region has a constant electrostatic potential due to its net electroneutrality, which means that bulk potential can be defined as a reference for the calculation of the specific electrode capacitance, $\Phi^{Bulk} = \Phi^{ref}$.[52] The resultant potential drop across the interface $\Delta\Phi = \Phi^{electrode} - \Phi^{ref}$, which is depicted in Figure 4, at a given surface charge density, $\sigma_s$, can be used to obtain the electrochemical double layer integral capacitance, $C_{EDL}$, of the electrochemical double layer,

$$C_{EDL} = \frac{\sigma_s}{\Delta\Phi - \Delta\Phi^0}. \quad\quad\quad 3$$

where $\Delta\Phi^0$ is the potential drop associated with the electrode at neutral charge.[53] The electrostatic potential profiles for all of the systems considered in this work are plotted in Sections S1 (QMMD) and S2 (MD) of the Supporting Information.

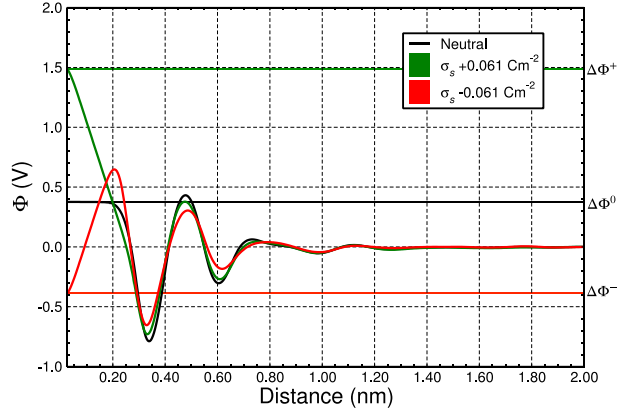

**Figure 4:** *Example plot for extracting potential drop from QMMD simulations at different surface charges. The electrostatic potential is plotted as a function of the distance from the electrode for the 1M LiCl system. Black, green and red solid horizontal lines mark the position where the potential drop of the neutral, positive and negative electrode is read.*

The treatment of the electrolyte permittivity for this family of interfaces has been subject to recent scrutiny.[54] Standard protocol is to approximate $\varepsilon_r$ using the dielectric constant of the electrolyte, yet it is well documented that truncation of the dipole field leads to a strong long-ranged dependence on the distance from the graphene surface. In fact, by fitting a polynomial function to the $\varepsilon_r(z)$ of SPC/E water from reference 55 Finney *et al* were able to extract the capacitance of graphite-NaCl solutions at various concentrations taking into account the changes in $\varepsilon_r$ due to proximity to the electrode.[54] For our systems, given that the electrostatic potential is calculated from a charge density constructed from all atomic and ionic partial charges, the different structuring of the water coordination shell shown in Figure 2b close to the electrode, are captured, thus we set $\varepsilon_r = 1$.

Another facet of the electrochemical interface we can explore with the QMMD method, which is not accessible by purely classical approaches, is the contribution of the electrode electronic structure to the integral capacitance. Where conventional metallic electrodes, which have near infinite DOS at the Fermi-level, effectively screening the contribution of the electrode electronic structure, a graphene electrode is two-dimensional and semimetallic. The ensuing finite density of states (DOS) close to the Fermi level responds to an applied potential by populating the conduction (valence) band edge with negative electrons (positive holes). The rate of change of the voltage of the electrode is different from the potential difference, and is referred to as the quantum capacitance, $C_Q$,[56] which contributes to the total specific electrode capacitance, $C_s$, as

$$\frac{1}{C_s} = \frac{1}{C_{\text{EDL}}} + \frac{1}{C_Q}. \qquad 4$$

The differential $C_Q$ can be directly obtained from the electrode DOS,

$$C_Q^{\text{diff}} = \frac{e^2}{4k_B T} \int_{-\infty}^{\infty} dE\, D(E) \operatorname{sech}^2\left(\frac{E + \Phi}{2k_B T}\right) \qquad 5$$

which in turn can be obtained at the DFTB-level of theory at each iteration of the QMMD loop. Here $E$ is the energy relative to the Fermi level, $D(E)$ is the DOS at a given energy, $e$ is the electron charge and $k_\text{B}T$ is the Boltzmann constant multiplied by temperature. Integration of the $C_\text{Q}^\text{diff}$ yields the integral quantum capacitance,[57]

$$C_Q = \frac{1}{\Phi}\int_0^\Phi d\Phi' C_\text{Q}^\text{diff}(\Phi').  \qquad 6$$

In practice, previous works have shown that the graphene DOS is resistant to changes in the presence of aqueous electrolytes,[58–60] as well as to charging of the sheet,[61,62] therefore a reasonable and commonly adopted approach is to use frozen bands;[63] the DFTB computed DOS and $C_Q$ in vacuum are plotted in Figure 5. We checked (see Supporting Information Section S5) how robust the electronic properties of the graphene are with respect to the various electrolytes and charged electrode states. To extract trends, we computed the running time averaged DFTB Fermi level (average over 250 frames) over 15 ns of each trajectory. We found that that the Fermi level expectedly shifts in response to charging, but that it is insensitive to the electrolyte composition. Moreover, as shown in Figure S6 of the Supporting Information, our simulations confirmed that the computed DOS is not just insensitive to the electrolyte, but also to the charge state, further validating the possible use of the frozen bands approximation. For each of the systems considered here, the computed $\Delta\Phi$, $C_\text{EDL}$ and $C_s$, are reported in Table 2.

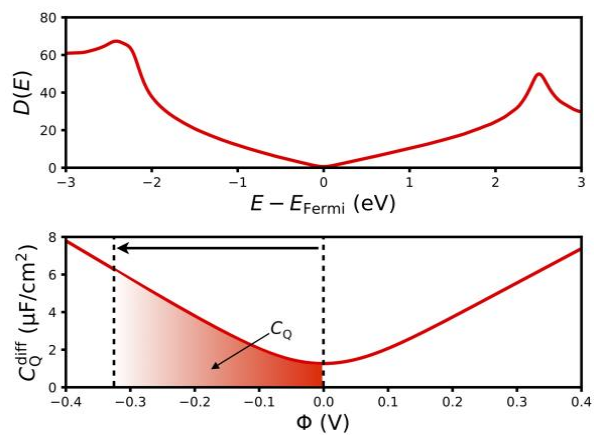

**Figure 5:** *Plots of the graphene electrode density of states (top) and the differential quantum capacitance ($C_\text{Q}^\text{diff}$) and area corresponding to the integral quantum capacitance ($C_Q$) (bottom). The solid arrow vertical dashed lines and solid arrows the voltage window and direction for which the quantum capacitance can be computed.*

**Table 2:** *Computed electrostatic potential drop ($\Delta\Phi$, reported in V), electrochemical double layer ($C_\text{EDL}$), quantum ($C_Q$) and specific ($C_S$) total capacitance (reported in µF cm$^{-2}$) at QMMD treated graphene cathode and anode charged neutral and to $\sigma_s = \pm 0.061$ Cm$^{-2}$. We also report the computed bulk dielectric constant for each electrolyte and the thickness of the double layer at the charged interface (reported in nm).*

|  | Neutral | -0.061 Cm$^{-2}$ | | | | +0.061 Cm$^{-2}$ | | | | | |
|---|---|---|---|---|---|---|---|---|---|---|---|
|  | $\Delta\Phi^0$ | $\Delta\Phi$ | $C_\text{EDL}$ | $C_Q$ | $C_s$ | $\Delta\Phi$ | $C_\text{EDL}$ | $C_Q$ | $C_s$ | $\varepsilon_r$ | $d_\text{EDL}$ |
| **LiCl** | 0.376 | -0.395 | 7.94 | 7.71 | 3.91 | 1.49 | 5.48 | 10.3 | 3.58 | 47.6 | 0.69 |
| **NaCl** | 0.385 | -0.359 | 8.22 | 7.42 | 3.90 | 1.52 | 5.41 | 10.4 | 3.56 | 48.9 | 0.74 |
| **KCl** | 0.356 | -0.365 | 8.48 | 7.18 | 3.89 | 1.50 | 5.36 | 10.5 | 3.55 | 49.4 | 0.75 |
| **MgCl$_2$** | 0.358 | -0.387 | 8.21 | 7.42 | 3.90 | 1.39 | 5.91 | 9.51 | 3.64 | 39.2 | 0.67 |
| **CaCl₂** | 0.311 | -0.408 | 8.51 | 7.15 | 3.89 | 1.41 | 5.58 | 10.1 | 3.60 | 39.9 | 0.71 |

In our setup, simulations are carried out at constant surface charge,[64] with the distribution of the surface charge density modulated through feedback with the electrolyte. It should be noted that direct comparison of computed capacitances of different electrode-electrolyte systems is difficult since it is not guaranteed that the electrode potentials are equivalent. For that reason we adopt the approach to Ho and Striolo to make such comparisons, as has been done in other works.[53,54,65] The computed $C_{\text{EDL}}$ and therefore $C_s$ depend on the measured potential drop according to equations 3 and 4. First we consider the capacitance of the electrochemical double layers of the various ions, which are reported in Table 2. At constant negative surface charge, the $C_{\text{EDL}}$ is roughly constant through the different monovalent ions, with a small increase observed moving from Li to K. The $C_{\text{EDL}}$ for the divalent ions are found to be equal in magnitude to the monovalent ions, with a small increase observed moving from Mg to Ca. On the other hand, at constant positive surface charge the $C_{\text{EDL}}$ of all systems are again found to be equivalent due to the independent adsorption of the Cl anion on the graphene in all cases (see Figure 1), with a small tendency to decrease moving from Li to K and Mg to Ca. Across all systems we find that the negatively biased electrode has a computed $C_{\text{EDL}}$ around 1.3 times greater than the comparative positively biased electrode. Finally, inspite of completely different structuring at the interface (e.g. 1M KCl), we also find that the computed $C_{\text{EDL}}$ values do not change between our QMMD and our MD simulations of the graphene electrode, see Table S1 of the Supporting Information.

To further understand the insensitivity of the electrochemical double layer capacitance to the different cations (as well as to the polarizability of the electrode surface) we have looked specifically at the ionic charge density profile across the electrochemical double layer. Integration of ionic charge density only, yields the concentration of excess charge (i.e. the balance between the cations and the Cl anions) within the electrochemical double layer. These excess charge concentrations (for the QMMD simulations) are reported in Table S2 of the Supporting Information. It is striking that irrespective of the structuring of the interface (KCl vs Li/NaCl), and irrespective of the cationic charge (KCl vs Mg/CaCl₂), the total concentration of excess charge in the electrochemical double layer is constant. In fact, in Figure S4 of the Supporting Information we compare explicitly as a function of distance from the electrode, the ionic charge density and excess charge concentration for the 1M KCl electrolyte at negatively charged QM and purely classical electrode. The result is that concentration of excess charge is equal approaching the end of the EDL. These results rule out local structuring of water as a compensating factor in the lack of specific ions effects on the $C_{\text{EDL}}$ value, and instead promote the idea that the electrochemical double layer capacitance is driven purely by the balance of cations to anions within the entire interfacial region.

Our computed values for the $C_{\text{EDL}}$ fall within the range 5-10 $\mu\text{F cm}^{-2}$, with the capacitance of the negative electrode being slightly larger than the positive electrode. This is in keeping with values reported in the literature for similar systems.[10,65] Previous MD simulations have observed a different trend for solutions of NaCl, KCl, RbCl and CsCl, where the $C_{\text{EDL}}$ decreased with ionic radius.[66] However, in reference [66] polarization effects are not accounted for and a different force field describing the pairwise non-bonded interactions was applied. A very recent contribution from Dockal *et al* reports essentially the opposite trend, with $C_{\text{EDL}}$ increasing with ionic radius;[65] in Dockal's work the polarizability of the surface is treated statically within the non-bonded Lennard-Jones parameters.[20] Yang *et al* addressed the dependence of $C_{\text{EDL}}$ on $\varepsilon_r$ in the Helmholtz layer by applying computed values of $\varepsilon_r$ for each electrolyte.[10] In their work, they computed a constant $C_{\text{EDL}}$ for the cation series Li to Cs, Ca and for mixtures of the cations. The constant $C_{EDL}$ is explained by the fact that the larger cations have both a larger dielectric constant and larger double layer thickness. According to the traditional Helmholtz theory, these two effects compete, resulting in a constant capacitance across the series, an effect we also observe in Figure 1. For our systems we obtained a constant double layer thickness and only marginally increasing dielectric constant for increasing ion sizes (see Table 2), moreover the divalent ions are found to have lower dielectric constant.

Concerning different electrodes, our results are different to those recently observed for a NaCl electrolyte in contact with gold electrodes, carried out using the constant-potential method to fix the electrode potential,[67] which found that a switch in absorption mechanism leads to different computed capacitances. The results presented here describe the situation of adsorption on a (semimetallic) graphene electrode. As we previously discussed, in the case of graphene the frozen bands approximation appears to be fully valid since because the electronic structure of the graphene is robust (Figure S6, Supporting Information). For metallic electrodes (such as Au[67]) this may not be the case, and inner Helmholtz layer ions adsorption could lead, through hybridization and modification of the electronic structure, to significant charging of the electrode and changes to the computed/measured capacitance.

We also tested the effect of doubling the concentration of the electrolyte on the computed values of the $C_{EDL}$ for the monovalent ions, these results are reported in Table 3, while a description of the systems and plots of the electrostatic potentials for each system are presented in the Supporting information, Section S1. We find that these systems are remarkably insensitive to the doubling of the ionic concentration. Similar effects have previously been observed for the measurement of the capacitance on other carbon based electrodes for example in the case of ionic liquids,[68] and linked to the net charge (difference between anions and cations) at the interface remaining constant at increasing concentration.

**Table 3:** *Computed electrostatic potential drop ($\Delta\Phi$, reported in V), electrochemical double layer ($C_{EDL}$), quantum ($C_Q$) and specific ($C_S$) total capacitance (reported in µF cm$^{-2}$) at QMMD treated graphene cathode and anode charged neutral and to $\sigma_s = \pm 0.061$ Cm$^{-2}$ for computed for the 2M electrolyte solutions.*

|  | Neutral | -0.061 Cm$^{-2}$ | | | | +0.061 Cm$^{-2}$ | | | |
|---|---|---|---|---|---|---|---|---|---|
|  | $\Delta\Phi^0$ | $\Delta\Phi$ | $C_{EDL}$ | $C_Q$ | $C_s$ | $\Delta\Phi$ | $C_{EDL}$ | $C_Q$ | $C_s$ |
| **LiCl** | 0.401 | -0.364 | 7.99 | 7.63 | 3.90 | 1.49 | 5.64 | 10.0 | 3.61 |
| **NaCl** | 0.366 | -0.367 | 8.34 | 7.31 | 3.89 | 1.49 | 5.42 | 10.5 | 3.56 |
| **KCl** | 0.313 | -0.406 | 8.50 | 7.12 | 3.89 | 1.45 | 5.38 | 10.5 | 3.56 |

3.3 Electronic Structure and Quantum Capacitance

Turning to the electrode electronic structure and computed quantum capacitance, besides a rigid shift in the absolute position of the Fermi energy, the DOS is found to be consistent across each of the different electrolytes and surface charge density regimes as shown in Section S5 of the Supporting Information. The fact that the energetic dispersion of the DOS does not change at different charge states indicates that the frozen bands approximation adopted for our simulations is valid. The computed DOS, reported in the top panel of Figure 5, shows that the DFTB approach captures the correct electronic features of graphene, namely the zero DOS and the linear dispersion of the band energies at the Fermi level. Integration of the DOS according to Equation 5 provides the differential quantum capacitance plotted in the bottom panel of Figure 5. The $C_Q^{diff}$ has a value of 2 µF cm$^{-2}$ at zero potential and increases with positive and negative bias. These results are in agreement with other approaches to obtain the $C_Q^{\text{diff}}$: for instance Xia *et al* employ a band theory model based on the two-dimensional electron gas (2DEG) and find that the minimum has a value of ~7 µF cm$^{-2}$.[69] Xia et al show that the 2DEG would have a zero $C_Q^{\text{diff}}$ without temperature induced broadening and impurity effects in the lattice. Interestingly, our results indicate only a minor asymmetry between the two branches of the differential capacitance, with the negative branch increasing slightly more rapidly, (Figure 5, bottom panel). On this basis, it can be said that strong asymmetries in the experimentally measured total specific capacitance arise from the $C_{\text{EDL}}$ component, i.e., preferential adsorption and structuring of either cations or anions on the surface. In other words, the Helmholtz capacitance for different ionic species cannot be considered constant.

Integration of $C_Q^{\text{diff}}$ over a potential window according to Equation 6, and represented pictorially in the bottom panel of Figure 5, yields the integral quantum capacitance. This can be combined with the EDL capacitance to calculate the total electrode specific capacitance as reported in Equation 4. As reported in Table 2, we find that the specific electrode capacitance (i.e. factoring the EDL and quantum effects) is constant with ionic radius moving from Li to K and with ionic charge from +1 to +2 on the negatively charged electrode surface. On the positively charged surface the specific capacitance is fairly constant across ionic radii and charge.

### 3.4 Diffusion of interfacial cations

Finally, we explore the dynamical properties of the ions at the interface. In comparison to the bulk liquid phase, water at the graphene-electrolyte interface was recently observed to have a higher self-diffusion,[42,70,71] this is the result of a redistribution of the hydrogen bonding network at the interface increasing the local packing of water molecules on the surface. In addition to this, other dynamical properties such as the water residence time at the surface changes when explicit polarizability of the surface and molecules are factored into the model.[72] Motivated by these changes in the behaviour of liquid water at the graphene interface, here we explore the ensuing changes to the self-diffusion coefficient for the different cations in contact with the charged polarizable graphene electrode. This is achieved by means of the Einstein Mean Squared Displacement (MSD),[72,73]

$$\lim_{\tau \to \infty} \langle \frac{\sum_i^N \left[ \left(x_i(t) - x_i(t+\tau)\right)^2 + \left(y_i(t) - y_i(t+\tau)\right)^2 \right]}{N 4 \tau P(\tau)} \rangle = D_C \qquad 4$$

modified to measure the displacement of particles in the plane of the electrode surface. The cation diffusion coefficient, $D_C$, is calculated from the slope of the MSD over time, where $P(\tau)$ is the particle survival probability, this is nothing more than the probability that a given cation will remain within a slab of thickness $\Delta z$ in the time interval $\tau$.

First we consider the self-diffusion of the water and ions in the bulk phase, these have previously been benchmarked for our simulation setup elsewhere[30,39] and help to validate our investigation on the interface. Far away from the electrode we recover the bulk diffusion coefficients of H$_2$O (0.224 Å$^2$ ps$^{-2}$), Li (0.083 Å$^2$ ps$^{-1}$), Na (0.103 Å$^2$ ps$^{-1}$), K (0.166 Å$^2$ ps$^{-1}$), Mg (0.047 Å$^2$ ps$^{-1}$) and Ca (0.058 Å$^2$ ps$^{-1}$) to within 0.01 Å$^2$ ps$^{-1}$. It should be noted that values for H$_2$O were extracted from the 1M KCl simulation. These results match trends observed for ion conductivities in the bulk: namely that conductivity, thence ion diffusion, increases from Li to K and decreases moving from an ion with oxidation state +1 to +2.[74]

For analysis of the interfacial diffusion, we partition the simulation box in to $\Delta z = 0.75$ nm slabs, this encompasses the local structuring of each ion at the surface. For each of the ions we find that the $D_C$ is larger near the surface, as reported in Table 4. On the other hand, water is found to have a lower $D_C$ (0.205 Å$^2$ ps$^{-1}$) in the EDL compared within the bulk. This is inconsistent with the purely classical picture for water (in contact with a single graphene sheet),[42,70] where the reorganization of the hydrogen bonding network into pentagonal rings at the surface results in a higher local packing of the water molecules compared to the bulk. Moreover, polarization effects are known to increase the residence time for water molecules on the surface, but previous results suggested that the diffusion coefficient is only minimally affected.[72] Therefore, we speculate that the presence of K ions at the surface in our simulations contributes to the disruption of the redistribution of the hydrogen bonded network,[42,70] combined with the polarizability of the electrode surface this leads to a decrease in the water lateral diffusivity. The increase to the diffusivity of the cations in the EDL can be linked to the position of their relative adsorption peaks in the density profiles presented in Figure 1b. The cations Li, Na, Mg and Ca all have their main adsorption peak "sandwiched" between the two main layers of water. These water layers therefore form a channel in the same plane as the electrode surface that the ions traverse with an increased in-plane $D_C$. It is striking that compared to the other cations K experiences a much less significant speed-up going from bulk to the interface. As noted in section 3.1 and as can be seen in Figure 1b, the K ions do not adsorb in the inter water layer spacing, rather directly within the two water layers resulting in a near constant $D_C$. When the electrode is charged, we see very little change to the self-diffusion of cations and water at the surface. This is not necessarily unexpected, very recent analysis of the H-bonding network of the water molecules seems show that the H-bonding network is only negligibly affected by the charge on the electrode.[65] In addition, Figure 1b shows that upon charging the cations (Li, Na, Mg and Ca) remain within the two water layers.

**Table 4:** *The computed in plane self-diffusion coefficients ($D_C$) of cations and H$_2$O at the surface (within 0.75 nm of the graphene) in a neutral and charged state and in the bulk.*

|  |  | $D_C$ (Å$^2$ ps$^{-1}$) | |
| --- | --- | --- | --- |
|  |  | Interface | Bulk |
| LiCl | 0 $e$ | 0.114 | 0.083 |
|  | 3.4 $e$ | 0.099 |  |
| NaCl | 0 $e$ | 0.122 | 0.103 |
|  | 3.4 $e$ | 0.125 |  |
| KCl | 0 $e$ | 0.175 | 0.166 |
|  | 3.4 $e$ | 0.143 |  |
| MgCl$_2$ | 0 $e$ | 0.069 | 0.047 |
|  | 3.4 $e$ | 0.065 |  |
| CaCl$_2$ | 0 $e$ | 0.077 | 0.058 |
|  | 3.4 $e$ | 0.079 |  |
| H$_2$O | 0 $e$ | 0.205 | 0.224 |
|  | 3.4 $e$ | 0.194 |  |

The changes to the cation $D_C$ moving from bulk towards the surface could also be the result of changes to the local concentrations of ions within EDL. The typical trend is that $D_C$ decreases with concentration,[30] while in our simulations the concentration of ions close to the surface (7.5 Å) is larger than in the bulk, as depicted in Figure 1. However, we can rule out the contribution of concentration since upon charging, the concentration of cations at the interface increases by at least a factor of 2, with no obvious change to their diffusion. That is, except for the K ion, which is significantly slowed with respect to the uncharged case. In the charged system, the concentration of K ions in the 1$^{st}$ layer (7.5 Å) is close to three times the uncharged case, and the number of ions that penetrate the inner Helmholtz layer is also greatly increased, Figure 1b. However, the change to $D_C$ is also likely a surface polarization effect: as we showed in Figure 3a and b, K ions in the inner Helmholtz layer strongly polarize the charged surface, this strong polarization-enhanced Coulomb interaction binds the ions to the surface atoms resulting in decreased in-plane mobility. The slowing of specific ions, based on their ability to bind to the surface, thence their hydration free energies, raise the interesting possibility for charged graphene electrodes, nanochannels and nanoflakes to act as capture devices for specific ions in the separation of mixed ionic solutions. We note, this dynamical effect cannot be observed in classical models that do not take explicit polarization into account, nor those which restrict fluctuations of the surface charge density to the locality of the atom since the requirement for the accumulation of charge at the adsorption site cannot be met.

We can also compare the dynamical behaviour of the ions at the interface to recent experiments on the conductivity of ions in nanochannels,[74] i.e. those with an interlayer separation equivalent in size to the hydrated ion. In those experiments, the mobility of Li and Na was found to increase moving from a bulk electrolyte to channel, K was found to have a roughly equivalent mobility while Mg and Ca both slowed. The change in behaviour is linked to the diameter of the hydrated ion, its hydration free energy and the ability for its hydration shell to distort or deplete. As we shown in Figure 2a and b, we find that the 1$^{st}$ solvation shell of the ion is unaffected by the presence of the surface except in the case of K. We also checked the 2$^{nd}$ solvation shell, which further confirms that only the solvation shell of the K is notably affected via adsorption on the surface. Therefore, taking into account our results and those of reference [74], we tentatively suggest that the increased mobility of the Li and Na observed in experiment is the result of exposure of the ion and its solvation shell to the graphene surface. The equivalence between the bulk and confined mobility of K results from a cancellation of the speed up due to its exposure to the surface and the direct adsorption of the ion polarizing either (or both) of the channel surfaces. Finally, the reduction in mobility of the Mg and Ca, which have both the largest ionic charges and solvated diameters, due to the forced dehydration of the ion entering the nanochannel and the highly charged ion strongly polarizing the surfaces and slowing down the diffusion dynamics. Clearly, whilst these insights provide promising explanations of the observed trends, further investigation, and validation of the behaviour of the ions within sub-nanometer cavities is required.

4. Conclusions

In summary, our QMMD simulations provide a new take on the contribution of specific ion effects to the thermodynamic and kinetic properties of aqueous graphene-electrolyte supercapacitors. We find that the coupling and interplay between dehydration and induced surface polarization dictates whether ions can penetrate the inner Helmholtz plane and directly adsorb on the surface. However, changes to the electrode specific capacitance with ionic radii are only accurate when the quantum capacitance is included. Regardless, complete discharge of the surface can be obtained by oppositely charging the electrode. Our calculations revealed that any asymmetries between the positive and negative branches of the measured capacitance originate in the EDL capacitance. This has implications for the modelling of a constant Helmholtz capacitance between solvated anions and cations. In the case of graphene, changes to the adsorption mechanism (inner vs outer Helmholtz adsorption) do not lead to changes in the measured electrochemical double layer capacitance, this can be linked to the robust electronic structure of the graphene electrode as well as to the balance of ionic charges over the entire double layer. Finally, we found that the diffusion of all ions at the interface is greater than their bulk values, while water does not change at the interface. The increase in the diffusivity of the ions can be linked to their adsorption in between the main adsorption layers of water suggesting the formation of ion channels in the EDL. K is the only ion in our study whose diffusivity is reduced upon charging of the electrode, this is due to the induced polarization of the graphene surface and highlights the important role surface effects have on ion mobilities on graphene, graphite and within graphitic nanochannels. The difference between diffusivity of K and the remaining ions Li, Na, Mg and Ca can only emerge in the case where the dynamics of water and ions are be treated on an equal footing. In this respect, where fully ab-initio methods can prove too costly, the lack of a solvation shell in implicit solvent models contributes to towards a failure to distinguish between ion adsorption in the inner and outer Helmholtz plane. Our results can inform on the future consideration of electrode-electrolyte couples and provide a basis for the extension of analytical treatment of the quantum mechanical polarization[51] to the cases of different ion pairs.

**Acknowledgements**

We acknowledge support provided by the IT Services use of the Computational Shared Facility (CSF) and at the University of Manchester. MC, JDE and PC thank the European Union's Horizon 2020 research and innovation programme project VIMMP under Grant Agreement No. 760907. AT acknowledges the support of the European Research Council (Grant No. 101020369).

**Author ORCIDS**

Joshua D. Elliott: 0000-0002-0729-246X
Mara Chiricotto: 0000-0003-1609-5254
Alessandro Troisi: 0000-0002-5447-5648
Paola Carbone: 0000-0001-9927-8376